# About Time: Observation of Time-Reflection at Optical Frequencies

**Ohad Segal[†1,2], Noa Konforty[†2,3], Oded Schiller[1,2], Maxwell J. Tolchin[5], Tony Yasniger[2,3], Michael Birk[2,4], Alexander Palatnik[2,3], Soham Saha[6], Pavel Sidorenko[1,2], Jon-Paul Maria[5], Yonatan Plotnik[2,3] and Mordechai Segev*[1,2,3]**

*1. Electrical & Computer Eng. Department, Technion-Israel Institute of Technology, Haifa, Israel.*
*2. Solid State Institute, Technion-Israel Institute of Technology, Haifa, Israel.*
*3. Physics Department, Technion-Israel Institute of Technology, Haifa, Israel.*
*4. The Norman Seiden Multidisciplinary Program in Nano-Science and Nano-Technology, Technion-Israel Institute of Technology, Haifa, Israel.*
*5. Department of Materials Science Engineering, Pennsylvania State University, USA.*
*6. Argonne National Laboratory, Lemont, IL 60439, USA*
*† These authors contributed equally to this work*
*msegev@technion.ac.il

**Abstract**:

Time-reflection occurs when a wave is propagating in a medium undergoing a large and abrupt change in its properties: the original wave splits into a time-refracted wave and a time-reflected wave, each displaying different features. The time-refracted wave continues along its original course but experiences a frequency shift, whereas the time-reflected wave is propagating backwards in space with a reversed phase, also with a shifted frequency. These phenomena are fundamental to any wave system, but the most interesting are electromagnetic (EM) waves, specifically at optical frequencies, where they can couple to light-matter interactions. However, time-reflection of EM waves was thus far observed only at RF frequencies, never at optical frequencies. This is because time-reflection requires an order-unity variation of the refractive index occurring faster than a single wave cycle, and conventional optical nonlinearities are either too weak or too slow by orders of magnitude. **Here, we present the first observation of time-reflection at optical frequencies.** We induce an order-unity refractive-index change with sub-cycle duration, observe the time-reflection, and study its fundamental properties. These results provide an experimental pathway to experimenting with time-interfaces, generating photonic time-crystals and exploring new regimes of light-matter interaction in time-varying media.

Waves in time-varying systems have captured the imagination of researchers from a variety of fields for over a century, ranging from Faraday waves[1] and the generation of quantum particles via the dynamic Casimir effect and Unruh acceleration[2–7] to general relativity effects in evolving systems[8] and the interaction between gravitational waves and light[9]. In the past decade, light in time-varying media and specifically in photonic time-crystals (PTCs) have drawn considerable attention. The fundamental building blocks of these phenomena, known as time-refractions and time-reflections, were first proposed in the 1950's[10,11]. When a wave propagates in a medium that is undergoing a large and abrupt change of its properties (a time-interface), the wave splits into two new waves: a time-refracted wave and a time-reflected wave. This scenario seemingly resembles what happens at spatial boundaries, but is in fact fundamentally different. Namely, instead of conserving energy (frequency), as for spatial interfaces, a time-interface in a homogeneous medium conserves momentum (wavevector). This means that the time-refracted and time-reflected waves experience a frequency shift at the time-interface. Moreover, because momentum is conserved, the time-refracted wave continues along its course (but at a shifted frequency), while the time-reflected wave is moving backward in space with conjugated phase – resembling a video playing backwards[12]. These time-refractions and time-reflections can interfere and give rise to exciting phenomena such as reciprocity breaking[13], inverse prisms[14], extreme energy transfer[15], temporal aiming[16], time-domain bound states in the continuum[17,18], non-resonant amplification[19], novel imaging possibilities[20] and photonic time-crystals[19,21–26].

Experimentally, the time-refracted waves are always present following a time-interface, regardless of the magnitude and timescale of the change in the properties of the medium. Indeed, time-refraction of waves has been observed in numerous physical systems, including EM waves at optical frequencies[27–31]. Time-reflections, on the other hand, only appear when the change in the material properties is both large (order-unity) and fast: faster than a single oscillation cycle of the wave propagating in the medium. For this reason, time-reflections have thus far been observed only with waves that have relatively low frequencies such as water waves[12], cold atoms[32] microwaves[33,34] acoustic[35] and elastic[36] waves and in synthetic dimensions[37,38], but never in the optical regime. Measuring a time-reflection at optical frequencies requires an order-unity change to the refractive index occurring within a few femtoseconds. Conventional (perturbative) nonlinear optics and other traditional mechanisms for changing the refractive index are not in this regime, because the change in the refractive index (or other property) is either too weak (e.g., the Optical Kerr effect or alike) or too slow (acousto-optic or photorefractive effects, etc.). This is the reason why time-reflection at optical frequencies has thus far never been observed.

**Here, we present the first experimental observation of time-reflection at optical frequencies.** To do that, we use a powerful few-femtosecond laser pulse to induce order-unity variation in the refractive index of Indium doped Cadmium Oxide (CdO) samples, and probe it with an infrared probe pulse for which the induced index change is in the deep sub-cycle regime. We study the time-reversal properties of the time-reflection and demonstrate that it retraces the trajectory of the incident probe pulse and that it preserves the polarization of a circularly-polarized probe beam, in sharp contradistinction from ordinary Fresnel reflection from a spatial interface which reverses the handedness. Moreover, by tuning the variation time of the refractive index from half-probe-cycle to multiple probe cycles, we measure the characteristic exponential dependence of the time-reflected wave on the variation time, as predicted by theory. Finally, we study the frequency translation of the time-reflected waves, arising from momentum conservation. These results provide an experimental pathway for exploring light-matter interactions in new regimes that have thus far been studied only theoretically, and for forming photonic time-crystals at optical frequencies, where they can couple to transitions between electronic energy levels.

Our experiments rely on extremely fast light-induced variations in the refractive index of materials known as TCOs (transparent conductive oxides). Over the past decade, these materials have sparked extensive experimental research on the evolution of light in time-varying media. On the technological side, TCOs were found to exhibit order-unity refractive index changes near their epsilon-near-zero (ENZ) point[27,28]. These index changes can be induced by ultrashort laser pulses enabling time-refractions induced by single cycle[29,30] and sub-cycle[31] order-unity index variations. On the theoretical side, band-structure analysis of materials with refractive index that is periodically modulated in time at order-unity contrast and few-femtoseconds period gives rise to the formation of PTCs[21,23,24]. When the refractive index is varied in such an abrupt fashion periodically in time, light undergoes multiple time-reflections and time-refractions that interfere with one another, giving rise to dispersion relations where continuous bands are separated by momentum gaps. The gap modes are special: when a pulse constructed from them enters the PTC, the pulse slows down and stops, while its amplitude grows exponentially by drawing energy from the modulation[24]. This exponential amplification of modes in PTCs was thus far observed only at RF frequencies[23] and in microwaves[39–41]. At optical frequencies specifically, where these EM phenomena can be coupled to transitions between electronic energy levels, such effects enable new technologies: non-resonant broadband light amplification, coherent light sources not relying on atomic resonances[19,25,42], new mechanisms of nonlinear frequency conversion[26,43] and generation of quantum light controlled by the temporal modulation[44]. However, as stated earlier, realizing time-reflections experimentally, and subsequently all the unique features displayed by time-varying media, requires materials with optical properties that can be significantly varied within a single cycle. For many years this was considered impossible at optical frequencies. In what follows we present the first experiments on time-reflections at optical frequencies.

Our time-reflection experiment is based on strongly varying the refractive index of a TCO material at a rate considerably faster than a single optical cycle. We do this by using powerful few-femtoseconds laser pulses at $0.8\mu m$ (mean) wavelength from a mode-locked Ti:Sapph laser, amplified to 7mJ pulse energy at 1KHz repetition rate, and subsequently compressed, using a gas-filled hollow core fiber (HCF), to 8fs FWHM carrying 0.5mJ energy. This laser pulse is the "modulator pulse": the pulse that induces the refractive index variation in our experiments. The sample we use is a $3\mu m$ thick layer of indium-doped CdO, with ENZ point around $4.3\mu m$ wavelength, deposited on sapphire. Accordingly, the “probe pulse”: the pulse that experiences the refractive index variation, is generated by passing part of the amplified Ti:Sapph laser beam through an optical parametric amplifier (OPA) followed by difference frequency generation (DFG), eventually resulting in a relatively long (120fs FWHM) probe pulse at $4\mu m$ mean wavelength and $0.5\mu J$ energy. A single oscillation of the EM field of the probe is of 13fs duration. The index variation is induced by optical excitation of charges, via two-photon absorption of the 8fs modulator pulse, from the valence band to the conduction band. Since absorption is a stimulated process, the increase in the population of conduction electrons and the resultant change in the refractive index follow the modulator pulse instantaneously. Accordingly, the index change occurs within ~60% of the probe cycle (see section S1 in [45]), which plays a critical role in generating time-reflection.

The experimental setup is sketched in Fig. 1. The modulator and probe pulses are synchronized to meet on the CdO sample with a controlled delay. The modulator pulse causes a sharp and large variation of the refractive index for wavelengths near the ENZ point of CdO, via two-photon absorption[31]. In our specific material and setting, the index variation is negative and of magnitude exceeding 0.6. However, to avoid confusion, henceforth we refer just to the magnitude of the variation in the index, ignoring the negative sign of the index change. Thus, in what follows the index variation is rising, and subsequently decays slowly over time. To maximize the spatial homogeneity of the refractive index variation experienced by the probe beam, we expand the modulator beam on the sample to ~2mm spot size, which is an order of magnitude larger than the diameter of the probe beam on the sample (~0.1mm). As a consequence of the large and extremely fast index change, the probe wave, traveling in the sample, splits into two new waves: a time-refracted wave and a time-reflected wave. As shown in Fig. 1, we measure the intensities of the time-reflected probe and the transmitted part of the probe (that includes the time-refracted part) that has passed through the sample, as well as the part of the probe that is Fresnel-reflected from the sample interface.

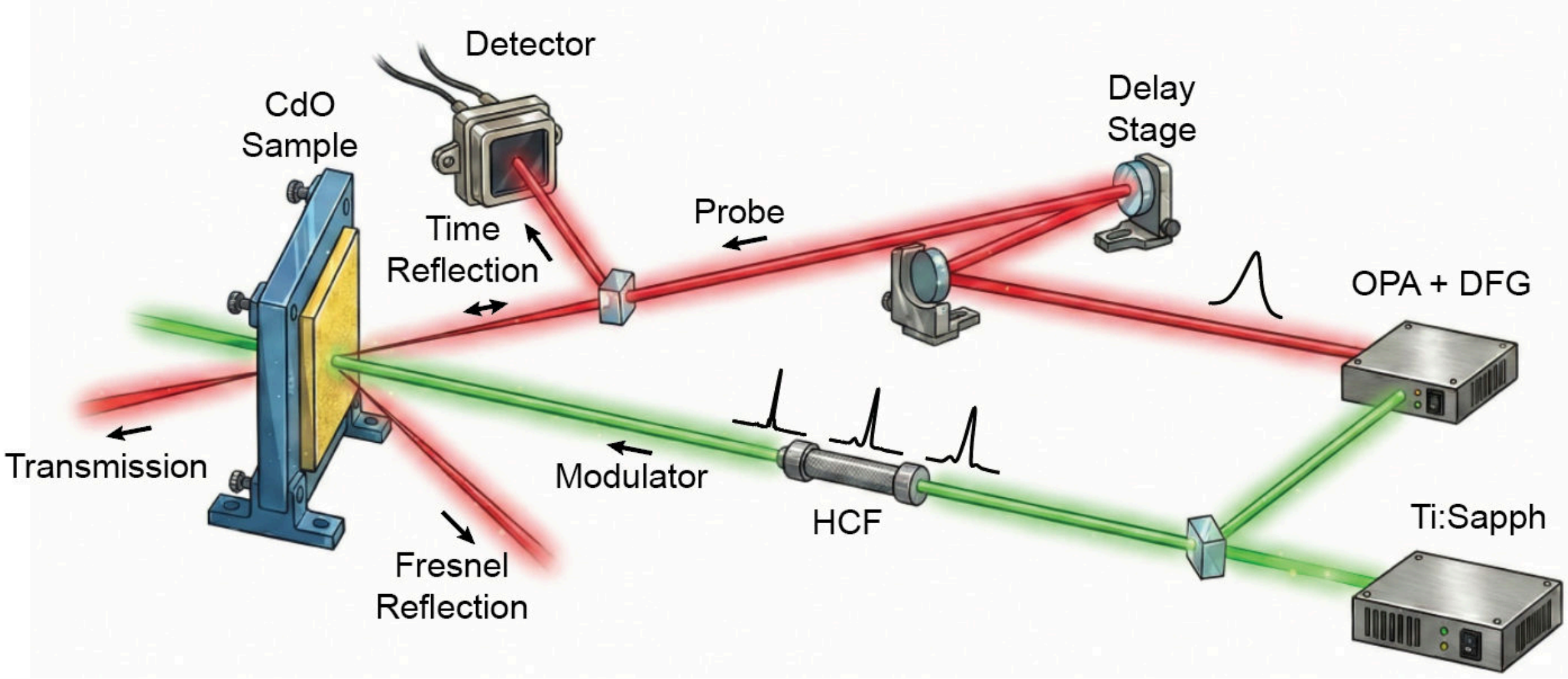

Figure 1: **Experimental setup for measuring time-reflections.** The modulator is a pulse at $0.8\mu m$ mean wavelength, compressed to various durations ranging from 8f to 40fs FWHM using an HCF. The probe is a 120fs FWHM pulse at $4\mu m$ mean wavelength (13fs cycle). The CdO sample has its ENZ point at $4.3\mu$m. The modulator changes the refractive index experienced by the probe by $\Delta n \approx -0.65$ within nearly half a cycle of the probe. This variation in the refractive index causes the probe pulse to be partially time-reflected, propagating counter to the propagation direction of the incident probe (see [45] section S5). Accordingly, the time-reflection is deflected using a beam-splitter and measured at the detector.

We study several features of the time-reflection in the experiment. The first is the trajectory of the probe. We intentionally launch the probe to be incident on the sample at a slight angle with respect to the normal to the surface. Since the time-reflected wave generated within the sample is propagating counter to the incident probe, it retraces the probe trajectory. Thus, after leaving the sample, we deflect the backward-propagating time-reflection by a beam-splitter into an extremely sensitive InSb detector (as illustrated in Fig. 1). In contradistinction, the Fresnel reflections of the probe from all the surfaces of the sample are reflected to a new trajectory, of opposite angle with respect to the normal. Consequently, the Fresnel reflections do not reach the InSb detector. The second fundamental feature of time-reversal is rooted in the polarization state of the probe pulse: we demonstrate the fundamental difference between the time-reflection and the Fresnel reflection of circularly-polarized light. Namely, the phase conjugate nature of the time-reflected waves dictates that the handedness of the circular polarization state in the lab frame is maintained, whereas any other wave reflected from an ordinary space-interface flips the handedness of its circular polarization state (see detailed calculation in [45] section S2). This is a special property of time-reflected waves, making them distinct from ordinary space reflections, as illustrated in Fig. 2b,c. We explore this property in our system as sketched in Fig, 2a.

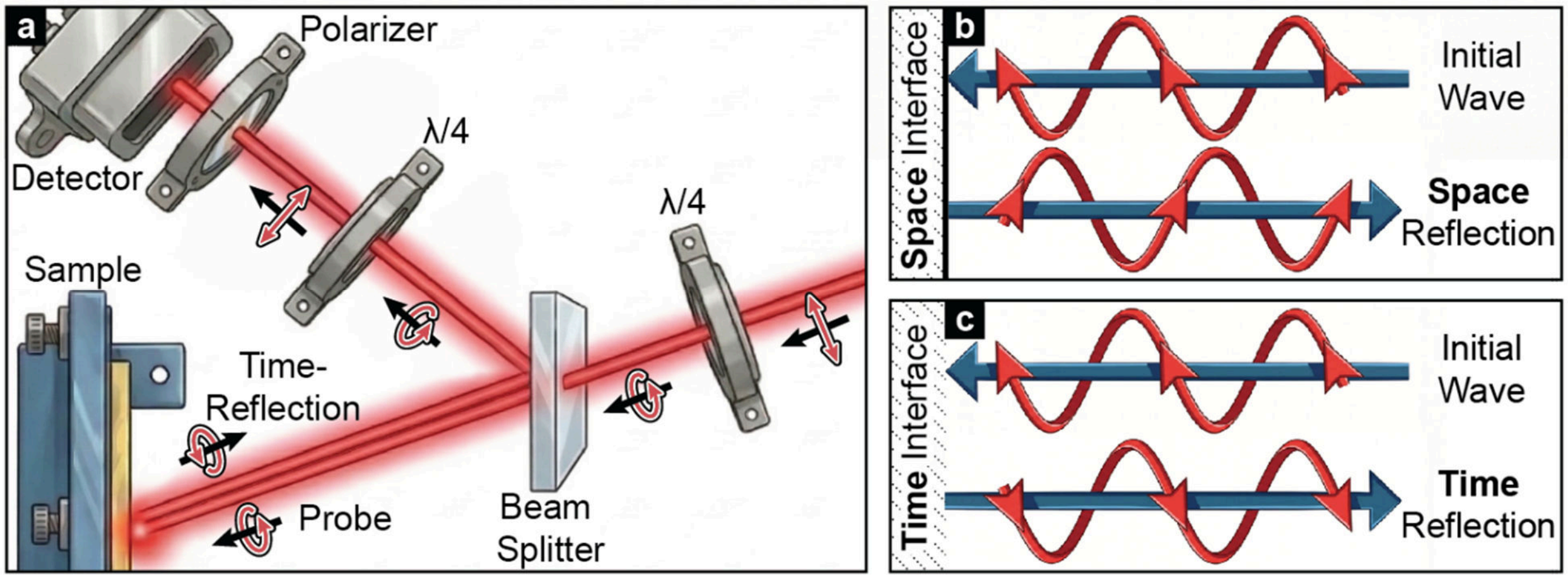

Figure 2: **Effects of space and time interfaces on circularly-polarized light. (a)** Experimental scheme for studying time-reflection of circularly-polarized light. The probe enters the scheme linearly polarized in the horizontal direction.

After passing through a quarter waveplate ($\lambda/4$) it changes its polarization to circular (clockwise rotating arrow). The circularly polarized probe passes through the beam-splitter and enters the sample. Inside the sample, part of it is time-reflected (TR) by the refractive index change. The TR is propagating counter to the original probe while maintaining the same circular polarization (reversed propagation direction and reversed rotating arrow, see [45] section S2) Subsequently, the TR reflects from the beam-splitter and passes through another quarter waveplate ($\lambda/4$) that changes its polarization back to linear horizontal. Just before the TR reaches the detector, it passes through a linear polarizer that transmits only linear horizontally polarized light (same polarization as the initial probe). This final polarizer isolates the TR from all Fresnel reflections and stray light in the system. **(b-c) Illustration comparing the effects of a space interface to a time interface on circularly-polarized light. (b)** Circularly polarized light (top red spiral arrow) is incident on a **space-interface** and reflected from it (bottom red spiral arrow). **(c)** Circularly polarized light is incident on a medium experiencing a **time-interface** and the resulting time-reversed time-reflection. The resulting circularly polarized time-reflected light is of opposite chirality from that caused by a spatial interface. This allows for isolation of the time-reflected wave from other reflections and stray light in the system.

The third feature of the time-reflection we study is the time-window within which it appears. As discussed above, time-reflection appears only if the probe is in the sample while the refractive index is varying significantly and at an extremely fast rate (faster than a single cycle). This is especially important in our experiment, since the temporal profile of the refractive index variation is a smooth step-like function (Fig. 3a). Namely, the magnitude of the index change is increasing following the modulator pulse, within a time-window shorter than 8fs. This index change is induced by two- photon absorption of the modulator pulse, which excites electrons from the valence band to the conduction band of the material, thereby changing the plasma frequency and leading to a negative change in the refractive index. This change is especially pronounced near the ENZ wavelength. Since the process is driven by optical excitation of electrons by the modulator pulse (absorption), the index change occurs within less than 8fs (Fig. 3a). In contrast, the relaxation of the index back to the original value involves recombination across the bandgap and is therefore slow (picoseconds time scale)[31]. This is why the index variation in our experiment is step-like, which has major implications: the time-reflection should appear only if (at least) part of the probe pulse is present in the sample within the time window of the 8fs while the index is varying extremely fast. On the other hand, the change in the Fresnel reflection and transmission (driven by the index change) should remain for many picoseconds after the passage of the modulator pulse, until eventually all the photoexcited charge carriers relax back to the valence band. It is thus informative to analyze the signals collected from the Fresnel reflection and the transmission of the probe as a function of modulator-probe delay, and examine the time-window within which the time-reflection appears, (Fig. 3). At negative delays before the -100fs mark in Figs. 3c-e, the probe arrives and exits the sample before the modulator pulse has arrived. During these early times, the probe simply passes through the sample and experiences ordinary Fresnel reflection and transmission from the spatial interfaces. In the other extreme – at positive delays beyond 100fs – probing the index change induced by the modulator after the modulator pulse has already left the sample - the probe pulse still experiences the long-lasting index change induced by the modulator because the index relaxation time is very long. Thus, at delays beyond 100fs, the difference in Fresnel reflection and transmission (with respect to the unmodulated values, before -100fs) is only due to the space interfaces of the sample and the new value of the refractive index, which is effectively static. Accordingly, as shown in Fig. 3c, the time-reflection signal is zero for delays before -100fs or after 100fs. ***The time-reflection signal appears only during the time window*** between [-100,100]fs, when the modulator is varying the refractive index while part of the probe pulse is present within the sample. Examining Fig. 3c emphasizes that the time-reflection signal is indeed present only within this time-window, when the modulator pulse is in the midst of changing the refractive index. This observation serves as another unique property of the time-reflection in our system: the slow relaxation of the index variation does not cause time-reflection.

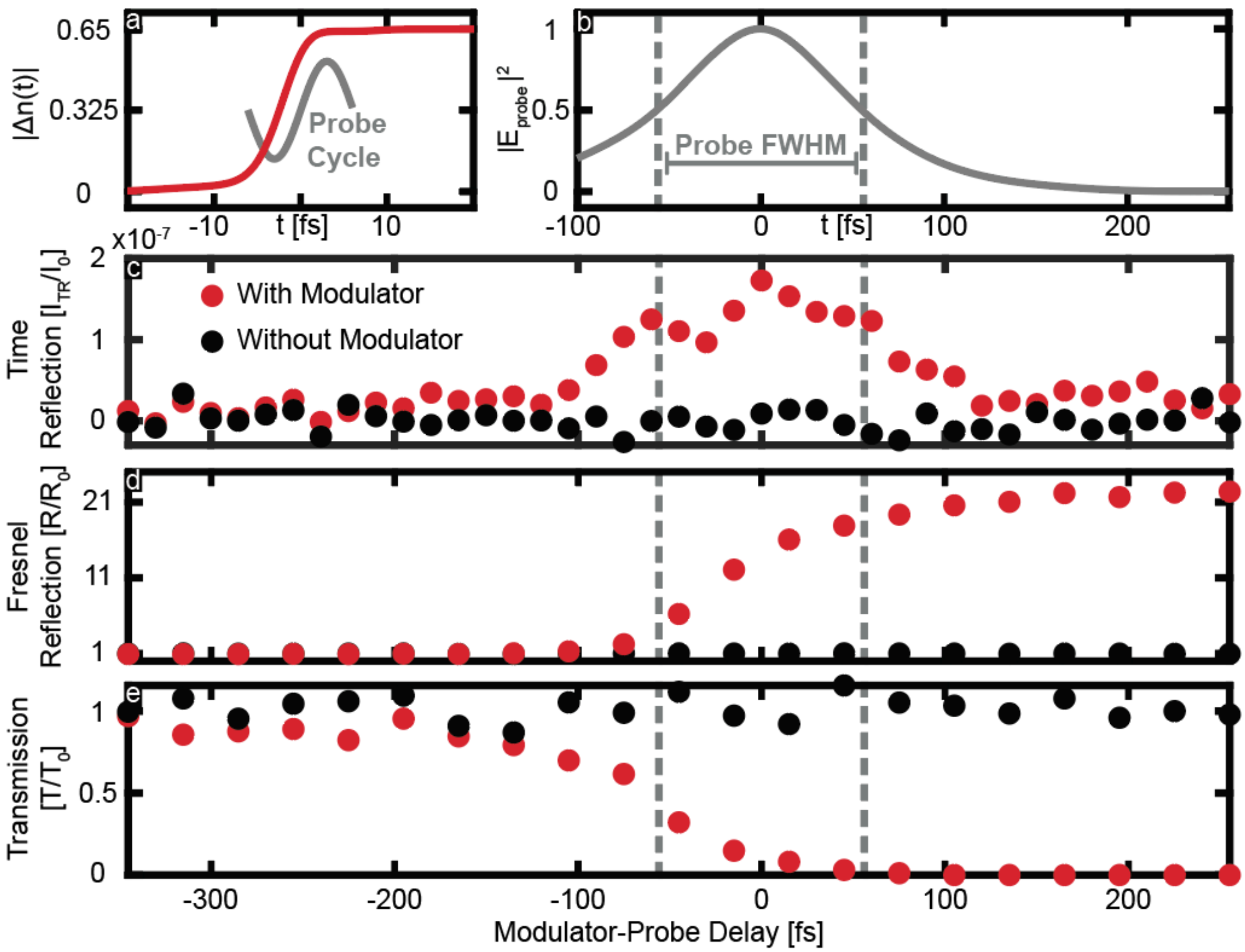

Figure 3: **Time-reflection, Fresnel reflection and transmission of the probe pulse with the 8fs modulator pulse.** Zero delay is set to the delay at which maximum time-reflection is measured. Dashed grey vertical lines show the FWHM of the incident probe pulse. **(a)** Step-like time-dependence of the magnitude of the refractive index change, calculated heuristically from the FROG reconstruction of the 8fs FWHM modulator pulse (see [45] section S1). For reference, a single cycle of the probe wave is shown in grey above the index step. **(b)** Intensity of the probe pulse reconstructed using XFROG. **(c)** Time-reflection of the probe pulse vs. modulator-probe delay. The time-reflection is generated in the sample, and it is propagating backwards retracing the trajectory of the incident probe wave with the correct handedness of the circular polarization. It appears only between $[-100,100]$fs delays, within which the refractive index is varying while the probe is in the sample. **(d)** Fresnel reflection of the probe pulse from the CdO sample interface vs. modulator-probe delay. Due to the refractive index variation, between delays $[-100,100]$fs, the Fresnel reflection is increasing. At longer delays, beyond 100fs, the Fresnel reflection remains enhanced due to the new value of the refractive index. **(e)** Transmission of the probe pulse through the CdO sample vs. modulator-probe delay. Due to the refractive index variation, between delays $[-100,100]$fs, the transmission is decreasing. At longer delays, beyond 100fs, the transmission remains low due to the new value of the refractive index.

Next, we examine another characteristic signature of time reflection in all wave systems: the strong dependence on the time scale during which the index change occurs. In fact, it is expected that the intensity of the time-reflection should decay exponentially when the rise-time of the index variation exceeds a single cycle of the probe wave. As explained in details in [45] section S4, this is a unique feature of time-reflection, as opposed to standard nonlinear processes which always scale with a power law of the time duration of the pump, never exponentially. Figure 4 shows this property of the time-reflected waves in our system. Using the HCF compression system, we stretch the modulator pulse from 8fs to 40fs and observe the exponential relation between the time-reflection intensity and the duration of the modulator pulse.

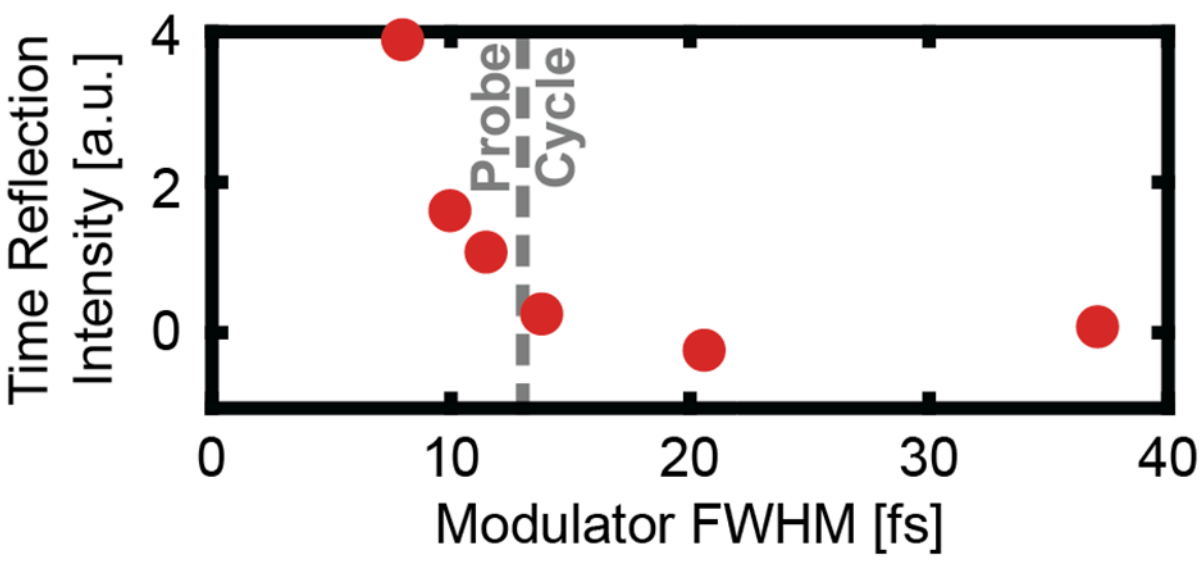

Figure 4: **Time-reflection power vs. modulator temporal width.** Stretching the duration of the modulator pulse from deep sub-probe-cycle to multi-probe-cycle reveals the exponential relation between the time scale of the refractive index variation and time-reflection intensity.

Last, recalling that the time-reflection should experience a frequency shift similar to the time-refraction, we measure the frequency shift of the time-reflected signal and compare to the spectra of the original probe pulse and of the time-refracted pulse. Because our sample is very thin ($3\mu m$) and only a fraction of the probe pulse is within the sample during the index variation, the intensity of the time-reflection in our experiment is ~$10^{-7}$ times weaker than our incident probe intensity. Our spectrometer is not sensitive enough to measure these time-reflections, and thus, to measure the frequency of the time-reflection, we use a bandpass filter. The bandpass filter (BP) is centered around $3.25\mu m$ wavelength (sufficiently away from the probe pulse) with a spectral transmission window of $0.5\mu m$. Accordingly, attempting to pass the original probe pulse through the BP filter, we find that the filter reduces the probe power by more than 97% (Fig. 5a). On the other hand, the BP filter reduces the time-reflection signal only by a factor of ~2 (Fig. 5b), implying that the spectrum of the time-reflection lies almost entirely within the high transmission window of the BP filter, which is strongly blue-shifted from the initial spectrum of the probe – as expected from the frequency shift of the time-reflection. Recall that the frequency shift of the time-reflection arises from conservation of momentum in homogeneous time-varying media and the lack of energy conservation. In our system, the induced index variation is negative (the overall refractive index is decreasing), thus the time-reflection (and the time-refraction) should be blue shifted with respect to the incident probe spectrum. This is exactly what we measure, as shown in Fig. 5.

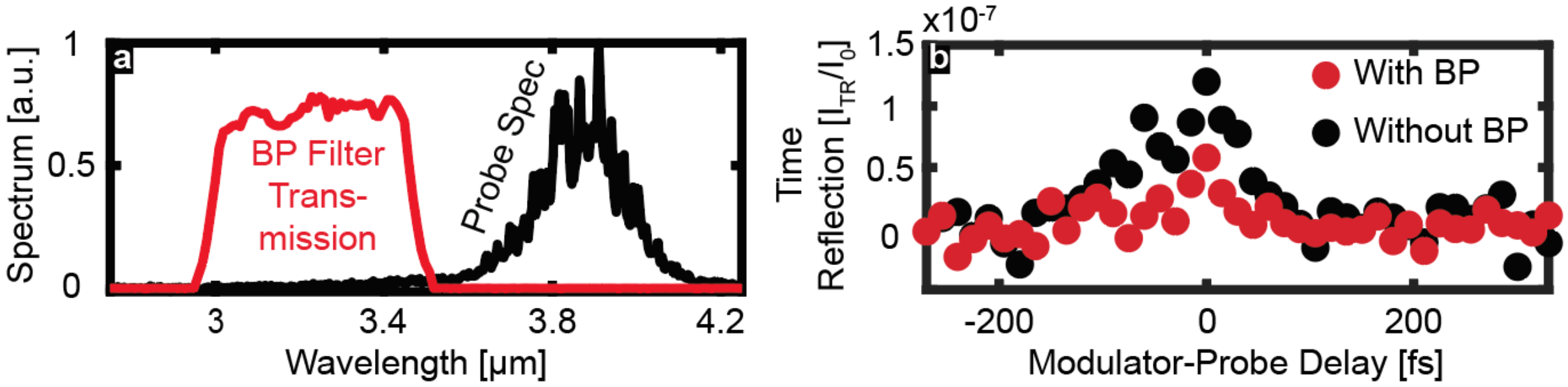

Figure 5: **Time-reflection spectrum measured using a band pass (BP) filter. (a)** Initial spectrum of the probe pulse (black line) measured using a spectrometer, and transmission function of the bandpass filter centered around $3.25\mu m$ wavelength with spectral width of $0.5\mu m$ (red line, adapted from[46]). The BP filter filters out more than 97% of the power of the original probe. **(b)** Time-reflection of the probe pulse vs. modulator-probe delay with (red data) and without (black data) the BP filter placed in front of the time-reflection detector. The power of the time-reflection is reduced only by factor 2, as expected from the blue-shift of the time-reflection.

Combining the 5 different features of time-reflection tested in our experiment, we can safely say that our results demonstrate the first observation of time-reflection and its attributes at optical frequencies. Specifically, the time-reflection exhibits the phase conjugation features manifested in propagation direction and polarization, the time-window of its appearance, the exponential scaling with the duration of the index variation, and the frequency shift. Observing time-reflection paves the way for the next step: cascading several time-interfaces and creating PTCs at optical frequencies. As we have shown here, our experimental system and the use of a TCO material with its ENZ point around $4\mu m$ wavelength, are suitable for generating PTCs. However, in order generate a PTC, we need both the rise and the relaxation of the index variation to generate time-reflection. The material used here has slow relaxation, because the relaxation mechanism is via recombination. This implies that we need an intra-band mechanism, as shown in[29]. A TCO material with a broader bandgap (or a modulator with a longer wavelength) and a similar ENZ point would be suitable. In such a medium, optical excitation by a modulator pulse would excite charges solely within the conduction band, hence the relaxation would also be in the single-cycle regime as observed in[29]. This will give rise to time-reflections and time-refractions during both the rise and the fall of the index change, and facilitate the generation of PTCs at optical frequencies. We envision that this will open the door for new technologies, ranging from widely tunable non-resonant amplification of light, coherent light sources not relying on atomic resonances[19,42], new mechanisms of nonlinear frequency conversion[26,43] and generation of quantum light controlled by the temporal modulation[44]. Time-reflections are the fundamental prerequisite for creating PTCs, hence our experimental demonstration serves as the critical gateway for unlocking the vision for a new realm of applications relying on photonic time-varying media.

This work was supported by the MAPATS program of the Israel Science Foundation and by the US Air Force Office of Scientific Research (AFOSR). OS gratefully acknowledges the support of the Adams Fellowships Program for excellent PhD students. The authors thank Gal Zeev for support with the graphical design of the figures.

## References

1. Faraday, M. On a peculiar class of acoustical figures; and on certain forms assumed by groups of particles upon vibrating elastic surfaces. *Proc. R. Soc.* 49–51 (1837) doi:10.1098/rspl.1830.0024.
2. Parker, L. Quantized Fields and Particle Creation in Expanding Universes. I. *Phys. Rev.* **183**, 1057–1068 (1969).
3. Moore, G. T. Quantum Theory of the Electromagnetic Field in a Variable-Length One-Dimensional Cavity. *J. Math. Phys.* **11**, 2679–2691 (1970).
4. Yablonovitch, E. Accelerating reference frame for electromagnetic waves in a rapidly growing plasma: Unruh-Davies-Fulling-DeWitt radiation and the nonadiabatic Casimir effect. *Phys Rev Lett* **62**, 1742–1745 (1989).
5. Dodonov, V. V., Klimov, A. B. & Nikonov, D. E. Quantum phenomena in nonstationary media. *Phys. Rev. A* **47**, 4422–4429 (1993).
6. Law, C. K. Effective Hamiltonian for the radiation in a cavity with a moving mirror and a time-varying dielectric medium. *Phys. Rev. A* **49**, 433–437 (1994).
7. Mendonça, J. T., Brodin, G. & Marklund, M. Vacuum effects in a vibrating cavity: Time refraction, dynamical Casimir effect, and effective Unruh acceleration. *Physics Letters A* **372**, 5621–5624 (2008).
8. Koivurova, M., Robson, C. W. & Ornigotti, M. Time-varying media, relativity, and the arrow of time. *Optica, OPTICA* **10**, 1398–1406 (2023).
9. Koufidis, S. Fr. & McCall, M. W. Coupling light waves to gravitational waves. *Phys. Rev. D* **112**, 104033 (2025).

10. Morgenthaler, F. R. Velocity Modulation of Electromagnetic Waves. *IRE Transactions on Microwave Theory and Techniques* **6**, 167–172 (1958).
11. Mendonça, J. T. & Shukla, P. K. Time Refraction and Time Reflection: Two Basic Concepts. *Phys. Scr.* **65**, 160 (2002).
12. Bacot, V., Labousse, M., Eddi, A., Fink, M. & Fort, E. Time reversal and holography with spacetime transformations. *Nature Phys* **12**, 972–977 (2016).
13. Sounas, D. L. & Alù, A. Non-reciprocal photonics based on time modulation. *Nature Photon* **11**, 774–783 (2017).
14. Akbarzadeh, A., Chamanara, N. & Caloz, C. Inverse prism based on temporal discontinuity and spatial dispersion. *Opt. Lett., OL* **43**, 3297–3300 (2018).
15. Li, H., Yin, S., Galiffi, E. & Alù, A. Temporal Parity-Time Symmetry for Extreme Energy Transformations. *Phys. Rev. Lett.* **127**, 153903 (2021).
16. Pacheco-Peña, V. & Engheta, N. Temporal aiming. *Light Sci Appl* **9**, 129 (2020).
17. Schiller, O., Plotnik, Y., Segal, O., Lyubarov, M. & Segev, M. Time-Domain Bound States in the Continuum. *Phys. Rev. Lett.* **133**, 263802 (2024).
18. Manzoor, Z., Schiller, O., Plotnik, Y., Segev, M. & Peroulis, D. Experimental Observation of Time-Domain Bound States in The Continuum. Preprint at https://doi.org/10.48550/arXiv.2604.10111 (2026).
19. Lyubarov, M. *et al.* Amplified emission and lasing in photonic time crystals. *Science* **377**, 425–428 (2022).
20. Schiller, O., Plotnik, Y., Bartal, G. & Segev, M. Negative Index Makes a Perfect Time-Domain Lens, Generating Slow Playback of Ultrafast Events. Preprint at https://doi.org/10.48550/arXiv.2512.03985 (2025).

21. Biancalana, F., Amann, A., Uskov, A. V. & O'Reilly, E. P. Dynamics of light propagation in spatiotemporal dielectric structures. *Phys. Rev. E* **75**, 046607 (2007).

22. Zurita-Sánchez, J. R., Halevi, P. & Cervantes-González, J. C. Reflection and transmission of a wave incident on a slab with a time-periodic dielectric function $ϵ(t)$. *Phys. Rev. A* **79**, 053821 (2009).

23. Reyes-Ayona, J. R. & Halevi, P. Observation of genuine wave vector (k or β) gap in a dynamic transmission line and temporal photonic crystals. *Applied Physics Letters* **107**, 074101 (2015).

24. Lustig, E., Sharabi, Y. & Segev, M. Topological aspects of photonic time crystals. *Optica, OPTICA* **5**, 1390–1395 (2018).

25. Dikopoltsev, A. *et al.* Light emission by free electrons in photonic time-crystals. *Proceedings of the National Academy of Sciences* **119**, e2119705119 (2022).

26. Konforty, N. *et al.* Second harmonic generation and nonlinear frequency conversion in photonic time-crystals. *Light Sci Appl* **14**, 152 (2025).

27. Shaltout, A. M. *et al.* Doppler-shift emulation using highly time-refracting TCO layer. in *2016 Conference on Lasers and Electro-Optics (CLEO)* 1–2 (2016).

28. Zhou, Y. *et al.* Broadband frequency translation through time refraction in an epsilon-near-zero material. *Nat Commun* **11**, 2180 (2020).

29. Lustig, E. *et al.* Time-refraction optics with single cycle modulation. *Nanophotonics* **12**, 2221–2230 (2023).

30. Tirole, R. *et al.* Double-slit time diffraction at optical frequencies. *Nat. Phys.* **19**, 999–1002 (2023).

31. Segal, O. *et al.* Sub-cycle time-refraction at optical frequencies. Preprint at https://doi.org/10.48550/arXiv.2601.05566 (2026).

32. Dong, Z. *et al.* Quantum time reflection and refraction of ultracold atoms. *Nat. Photon.* **18**, 68–73 (2024).

33. Moussa, H. *et al.* Observation of temporal reflection and broadband frequency translation at photonic time interfaces. *Nat. Phys.* **19**, 863–868 (2023).

34. Jones, T. R., Kildishev, A. V., Segev, M. & Peroulis, D. Time-reflection of microwaves by a fast optically-controlled time-boundary. *Nat Commun* **15**, 6786 (2024).

35. Kim, B. L., Chong, C. & Daraio, C. Temporal Refraction in an Acoustic Phononic Lattice. *Phys. Rev. Lett.* **133**, 077201 (2024).

36. Wang, S. *et al.* Experimental realization of temporal refraction and reflection in elastic beams. *Nat Commun* **16**, 9520 (2025).

37. Feis, J., Weidemann, S., Sheppard, T., Price, H. M. & Szameit, A. Space-time-topological events in photonic quantum walks. *Nat. Photon.* **19**, 518–525 (2025).

38. Palatnik, A. *et al.* Time-reflection in a discrete synthetic photonic lattice. in *CLEO 2024 (2024), paper FTh1L.5* FTh1L.5 (Optica Publishing Group, 2024). doi:10.1364/CLEO_FS.2024.FTh1L.5.

39. Xiong, J. *et al.* Observation of wave amplification and temporal topological state in a non-synthetic photonic time crystal. *Nat Commun* **16**, 11182 (2025).

40. Jones, T. R. *et al.* Observation of Photonic Time-Crystals at microwave frequencies. *submitted*.

41. Huang, L. *et al.* Microwave vortex beam lasing via photonic time crystals. Preprint at https://doi.org/10.48550/arXiv.2601.00585 (2026).

42. Tulchinsky, M., Plotnik, Y., Schiller, O. M., Konforty, N. & Segev, M. Towards a Photonic Time Crystal laser: Modeling Saturation and Loss. in FS5.

43. Saha, S. *et al.* Third Harmonic Enhancement Harnessing Photoexcitation Unveils New Nonlinearities in Zinc Oxide. Preprint at https://doi.org/10.48550/arXiv.2405.04891 (2024).

44. Lyubarov, M. *et al.* Continuous-variables cluster states in photonic time-crystals. *Optica Quantum, OPTICAQ* **3**, 366–371 (2025).

45. See Supplemental Material at [URL] for more details, calculations and parameters, which includes Refs.

46. Thorlabs · FB3250-500 Ø1" IR Bandpass Filter, CWL = 3.25 µm, FWHM = 500 nm. https://www.thorlabs.com/item/FB3250-500 https://media.thorlabs.com/contentassets/5d43115faba44241bf9e6cfb79191e1b/fb3250-500.xlsx?v=1202104802.